\documentclass[10pt,a4paper]{fullarticle}
\usepackage[american]{babel}
\usepackage{csquotes}
\usepackage{sciencestuff}

\input{arxiv.def}
\usepackage{easyReview}
\setreviewson

\hypersetup{%
    pdftitle={Field Theory of Data: Anomaly Detection via the Functional Renormalization Group},
    pdfauthor={Riccardo Finotello, Vincent Lahoche, Parham Radpay, Dine Ousmane-Samary},
    pdfsubject={Functional Renormalization Group},
    pdfkeywords={renormalization group, field theory, random matrix theory, signal analysis, information theory},
}

\title{%
    Field Theory of Data: Anomaly Detection via the Functional Renormalization Group \protect \\
    {\Large The 2D Ising Model as a Benchmark}
}

\author[1]{Riccardo Finotello\emailfoot{riccardo.finotello@cea.fr}}
\author[2]{Vincent Lahoche\emailfoot{vincent.lahoche@cea.fr}}
\author[2]{Parham Radpay\emailfoot{parhamradpay@gmail.com}}
\author[3]{Dine Ousmane-Samary\emailfoot{dine.ousmanesamary@cipma.uac.bj}}
\affil[1]{%
    Université Paris-Saclay, CEA,
    \protect \\
    Service de Génie Logiciel et de Simulation (SGLS),
    \protect \\
    Gif-sur-Yvette, F-91191, France
}
\affil[2]{%
    Université Paris-Saclay, CEA,
    \protect \\
    Palaiseau, F-91120, France
}
\affil[3]{%
    Faculté des Sciences et Techniques (ICMPA-UNESCO Chair), Université d'Abomey-Calavi,
    \protect \\
    Cotonou, BP 526, Benin
}

\date{}

\renewcommand{\dd}{\mathrm{d}}
\newtheorem{theorem}{Theorem}
\newtheorem{definition}{Definition}

\begin{document}

\maketitle

\begin{abstract}
    We establish a correspondence between anomaly detection in high-noise regimes and the renormalization group flow of non-equilibrium field theories.
    We provide a physical grounding for this framework by proving that the detection of phase transitions in interacting non-equilibrium systems maps to the study of an effective equilibrium field theory near its Gaussian fixed point, which we identify with the universal Marchenko-Pastur distribution.
    Applying the Functional Renormalization Group to the two-dimensional Model A, we demonstrate that the noise-to-signal ratio acts as a physical temperature, where the signal emerges as ordered domains within a thermalized background of fluctuations.
    Using the exact Onsager solution as a benchmark, we show that this approach identifies critical thresholds with an error below 4\%, significantly outperforming standard information-theoretic metrics such as the Kullback-Leibler divergence.
    Our results provide a universal strategy for resolving structures in complex datasets near criticality, bridging the gap between statistical mechanics and statistical inference.
\end{abstract}

\highlights{The functional renormalization group is used to identify a known non-equilibrium phase transition as a benchmark for data science applications.}

\clearpage

\tableofcontents

\clearpage

\section{Introduction}\label{sec:introduction}

In this article, we investigate the correspondence between signal detection methods based on field theory and the Functional Renormalization Group (FRG), and non-equilibrium critical phenomena.
The application of the FRG to signal detection in data science has matured through a series of methodological developments focused on the flow of empirical spectra~\cite{RG0,RG1,RG2,RG3,RG4,RG5,RG6,RG7}.
This work establishes a rigorous connection with the phenomenology of non-equilibrium statistical mechanics by grounding the renormalization group framework for signal detection in the non-equilibrium dynamics of Model A.
We construct our effective field theory using the Maximum Entropy Principle, which provides the least biased representation of the data.
We demonstrate that signal detection, interpreted as a phase transition in the emergent effective field theory, can be validated by its ability to resolve physical phase transitions.
The bridge is motivated by the coarsening regime of the 2D non-equilibrium Ising model, where spin domains emerging after a quench exhibit multi-scale correlations~\cite{livi2017nonequilibrium}.
Such structures are characteristic of real-world datasets, such as images, where pixels remain correlated over finite domains.
In this framework, the temperature $T$ governs the emergence of these domains and serves as a physical analog to the Noise-to-Signal Ratio (NSR).
High temperatures correspond to noise-dominated regimes in which stochastic fluctuations suppress the formation of ordered structures.
The spontaneous formation of ordered domains, characterized by the dynamics of domain walls, represents the emergence of non-trivial correlations within the stochastic background.
Critical behavior is characterized by the competition between pixel ordering and power-law fluctuations encoded in the correlation matrix~\cite{vinayak2014spectral}.
This work relates the methodology developed in~\cite{RG0,RG1,RG2,RG3,RG4,RG5,RG6,RG7} with signal detection in high-NSR regimes, where traditional analysis is most constrained.

\begin{figure}[t]
    \centering
    \begin{tikzpicture}[scale=1.3]
    \draw[draw=red, fill=red!25, pattern=north east lines, pattern color=red, domain=0.5:4.0, samples=250, smooth] plot (\x,{sqrt(4.0 - \x)*sqrt(\x - 0.5)/\x});

    \draw[thick, dashed, draw=black, fill=black!10, domain=0.25:3.5, samples=250, smooth] plot (\x,{sqrt(3.5 - \x)*sqrt(\x - 0.25)/(0.85*\x)});
    \node[align=center] at (1.25, 0.5) {bulk};

    \draw[dash dot, black] (3.55,-0.15) node[below] {$\Lambda$} -- (3.55,2.5);
    \draw[dash dot, red] (4.05,-0.15) node[below] {$\Lambda^{\prime}$} -- (4.05,2.5);

    \draw[black] (4.2,0) -- (4.2,0.8);
    \draw[black] (4.3,0) -- (4.3,0.9);
    \draw[black] (4.5,0) -- (4.5,0.65);
    \draw[black] (4.8,0) -- (4.8,1.0);

    \draw[->, thick] (-0.15,0) -- (5,0) node[above] {$\lambda$};
    \draw[->, thick] (0,-0.15) -- (0,2.75) node[left] {$\mu$};
\end{tikzpicture}
    \caption{%
        The presence of a heavy tail (red pattern) in the spectrum is usually hard to separate from the true noise (the \emph{bulk}).
        While isolated spikes can be identified as the principal components of the data, a noise model (black dashed line) derived from random matrix theory provides a reliable separation.
    }\label{fig1}
\end{figure}
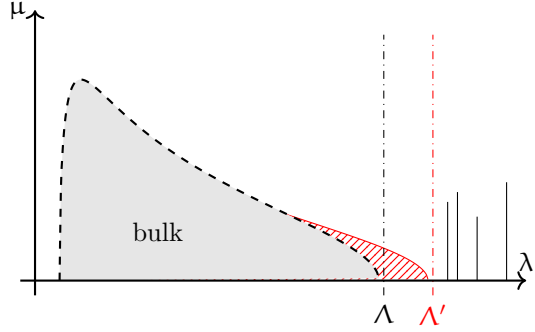

In data science, raw data often take the form of large matrices $X \in \mathbb{R}^{N \times P}$, where $N$ is the sample size (e.g., the number of images in the dataset) and $P$ represents the dimensionality of the data (e.g., the number of pixels in the images).
The correlation matrix $C \in \mathbb{R}^{P \times P}$ encodes the minimal information required for any signal analysis.
Its entries are defined by:
\begin{equation*}
    C_{ij}=\frac{Q_{ij}}{\sqrt{Q_{ii} Q_{jj}}},
\end{equation*}
where $Q = (X^T X) / N$ is the \emph{covariance matrix}.
Since the data may contain both signal and noise, $C$ also contains a random component, generally associated with the latter.
The primary objective of analysis techniques is to isolate these noisy degrees of freedom.
In favorable cases (low NSR), the spectrum of $C$ exhibits a quasi-continuous bulk, with eigenvalue spacing of order $O(N^{-1})$, alongside a number of isolated \emph{spikes} (Figure~\ref{fig1}, left).
In this configuration, the boundary between noise and true information is clear: it is defined by the bulk edge, which often corresponds to a universality class of Random Matrix Theory (RMT).
Signals associated with these spikes can then be recovered using standard, computationally efficient methods such as Principal Component Analysis (PCA)~\cite{PCAAPP}.
Conversely, at high NSR, the quasi-continuous spectrum may spread (heavy-tail effect), making the definition of the boundary between noise and signal ambiguous (Figure~\ref{fig1}, center).
RMT provides an initial solution.
Theorems regarding the spectra of large random matrices enable the identification of noise-dominated regions within the spectrum.
Notably, the Marchenko-Pastur (MP) theorem~\cite{Bouchaud3} describes the spectrum of an infinite-dimensional Gaussian correlation matrix:

\begin{theorem}\label{th1}
    Let $X \in \mathbb{R}^{R \times S}$ be a matrix with i.i.d.\ entries and variance $\sigma^2$.
    As $R, S\to \infty$ with the ratio $q = R/S$ fixed, the empirical eigenvalues distribution of the corresponding $S \times S$ random Wishart matrix $Z= X^T X / R$ converges weakly toward the MP distribution:
    \begin{equation}
        \mu_{MP}(\lambda)
        =
        \frac{\sqrt{(\lambda_+-\lambda)(\lambda-\lambda_-)}}{2\pi \sigma^2 q \lambda},
        \label{MP}
    \end{equation}
    where $\lambda_\pm = \sigma^2 (1 \pm \sqrt{q})^2$.
\end{theorem}

The identification of the ``signal-noise'' boundary using a noise model inspired by random matrix theory (Figure~\ref{fig1}, right) was historically introduced by Bouchaud and Potters~\cite{Bouchaud1} for the analysis of financial markets and was subsequently further developed~\cite{Bouchaud2,bun2017cleaning,Utsugi,Plerou,Guhr,Burda,Akemann2}.
Within this framework, the spectrum plays the role of the momentum distribution in the statistical field theory, which determines the relevant or irrelevant nature of interactions through a power-counting scheme that is uniquely scale-dependent.
A review of the core principles of this method, hereafter referred to as Generalized Scale Analysis (GSA), is provided in Section~\ref{SDRG}.
Additionally, the reader may consult the review~\cite{RG5}, as well as the more recent paper~\cite{RG7}.

This work presents two primary contributions.
First, the FRG is used to resolve signal detection problems with higher precision than traditional information-theoretic techniques.
Second, we prove that identifying phase transitions in interacting non-equilibrium systems can be mapped to the study of an effective equilibrium field theory near its Gaussian fixed point.
By benchmarking this approach against the exact Onsager solution, we demonstrate that the FRG captures critical thresholds with higher accuracy than standard information-theoretic metrics.
This physical grounding establishes the FRG as a robust tool in the data scientist toolbox for detecting changes in underlying regular behavior by resolving the critical region of a non-equilibrium phase transition.

While the core phenomenology of non-equilibrium phase transitions is established for simple models, the precise calculation of critical exponents and the study of complex systems, such as spin glasses, quantum transitions, and active matter, remain active areas of research~\cite{keller2016active}.
The GSA framework provides a numerical strategy for estimating physical observables in these contexts.
Furthermore, the application of the FRG to signal detection offers new possibilities for analyzing complex datasets, especially in high-NSR regimes where traditional methods are often insufficient.

\section{Model A and its numerical implementation}\label{ModelA}

Model A is a kinetic Ising model that appears in the coarse-graining limit of Glauber dynamics for Ising spins.
In $D$ dimensions, discrete spins are expressed after coarse graining by a \emph{continuous field} $\varphi(x,t)$ following Langevin-type dynamics~\cite{Zinn1,le2004equilibrium,livi2017nonequilibrium}.
For simplicity, we set the Boltzmann constant to unity $k_B=1$:
\begin{equation}
    \frac{\partial}{\partial t} \varphi(x,t)
    =
    - \frac{1}{T} \frac{\delta H}{\delta \varphi(x,t)}+ \eta (x,t),
    \label{eqLangevin}
\end{equation}
where $T$ is the temperature of the equilibrium system and $H$ is the time-dependent Ginzburg-Landau Hamiltonian:
\begin{equation}
    H =\int \dd x \dd t\, \left(\frac{1}{2} (\nabla \varphi)^2+V(\varphi) \right).
\end{equation}
The potential $V$ is assumed to be a polynomial $V(\varphi) = \frac{1}{2} a \varphi^2+ \frac{b}{4} \varphi^4+ \cdots$.
The term $\eta(x,t)$ is white Gaussian noise with zero mean\footnote{The factor $2$ ensures that the equilibrium probability distribution matches the Boltzmann distribution $P(\varphi) \sim e^{-H/T}$, see~\cite{Zinn1}.}:
\begin{equation}
    \langle \eta(x,t) \eta(x^\prime,t^\prime) \rangle_\eta = 2 \delta(x-x^\prime) \delta(t-t^\prime).
\end{equation}
Equation~\eqref{eqLangevin} represents a stochastic gradient descent toward a minimum of the Hamiltonian.
In this article, we simulate a 2D model based on the following discretization:
\begin{equation}
    \varphi_{i,j}^{n+1}
    =
    \varphi_{i,j}^{n} + \Delta t \left[ \frac{1}{T} (\Delta_{\text{dis}}\varphi)^{n}_{i,j} - V^{\prime}[\varphi^{n}]\right]+\sqrt{2\Delta t}\,\xi_{i,j}^{\,n}
    \label{discrete}
\end{equation}
where $\xi_{i,j}^{\,n}\sim\mathcal{N}(0,1)$ is spatiotemporally uncorrelated. The discrete Laplacian is:
\begin{equation}
    (\Delta_{\text{dis}}\varphi)_{i,j}
    =
    \frac{\varphi_{i+1,j} + \varphi_{i-1,j}+\varphi_{i,j+1} + \varphi_{i,j-1}-4\varphi_{i,j}}{ \ell^2}.
    \label{DiscreteLaplacian}
\end{equation}
The hyperparameter $\ell$ corresponds to the lattice spacing and is fixed at $1$.
Furthermore, we choose the following quartic potential:
\begin{equation}
    V(\varphi) = \frac{1-T^{-1}}{2}\,\varphi^2 + \frac{b}{4T}\varphi^4.
\end{equation}
This form has a numerical advantage over the standard $(T-T_0)$ form, to which it is equivalent near the transition, by avoiding an excessively large mass term.
The bare transition temperature $T_0$ is fixed at $1$.
When $T<1$, the potential exhibits a double-well structure and $\varphi=0$ becomes an unstable solution.
However, $T_0$ is not the physical critical temperature $T_c$.
Due to the quartic interaction, the system displays an additional rigidity from fluctuations that corrects the bare temperature value (see Appendix~\ref{sec:app1})~\cite{Hoenberg1977}.
The code of the simulation can be found on \href{https://github.com/ParhamRadpay/Model-A}{https://github.com/ParhamRadpay/Model-A}.

\begin{figure*}[t] 
    \centering
    \includegraphics[width=0.28\textwidth]{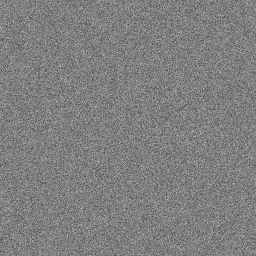} \hfil
    \includegraphics[width=0.28\textwidth]{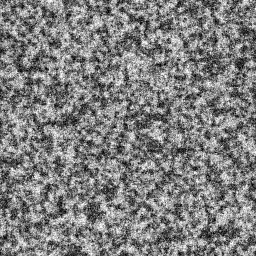} \hfil
    \includegraphics[width=0.28\textwidth]{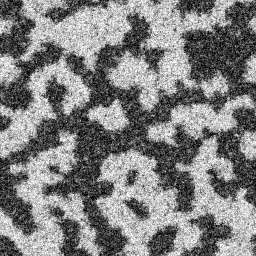}
    \caption{%
        Numerical simulation of a quench for $N = 100$: starting at $t=0$ at a very high temperature, the spins are plunged into a thermal bath below the critical temperature. The left image shows step 0, and the next two images correspond to time steps $3 \times 10^3$ and $39 \times 10^3$.
    }\label{fig2}
\end{figure*}

The correspondence with the 2D Ising model can be established heuristically as follows.
For simplicity, we consider a unit Boltzmann constant.
The critical temperature in mean-field theory is $T_0=4J$, where $J>0$ is the Ising coupling in the Hamiltonian $H_\text{Ising} = J \sum_{\langle i,j \rangle} S_i S_j$, and $\langle i,j \rangle$ denotes the sum over nearest neighbors.
For $D=2$, Onsager calculated $T_c \approx 2.27 J$.
Setting $T_0=1$ fixes $J=1/4$, which leads to $T_c \approx 0.57$.
Moreover, the formal correspondence at the critical point requires $b = 1/(3 T_c^3) \approx 1.80$.
This value of $b$ is determined by a fourth-order truncated Ginzburg-Landau expansion derived from a Hubbard-Stratonovich transformation.
Although this mapping is exact at the level of the effective action, numerical convergence toward the Onsager temperature is only achieved in the limit of an infinitely deep potential well ($b \to \infty$).
Numerical coincidences for finite values of $b$ correspond to Ising-like systems.
A more robust correspondence with the standard 2D Ising model can be established by considering the potential (see Appendix~\ref{sec:app2})
\begin{equation}
    V(\varphi)=-\frac{1}{2} \varepsilon\varphi^2+\frac{1}{4}\varepsilon \varphi^4
    \label{VIsing}
\end{equation}
with the limit $\varepsilon \to \infty$, which enforces the constraint $\varphi=\pm 1$.
The choice $J=1/4$ then requires scaling the Laplacian in~\eqref{discrete} by a factor of $1/4$.
In the following, the model described by~\eqref{VIsing} will be referred to as the Ising model.
The simulation is performed on a grid of size $N$, typically $N = 100$, and tracks the dynamics over $t_{\text{fin}} = 4 \times 10^4$ steps.
The data are presented as a large matrix $X \in \mathbb{R}^{N^2\times t_{\text{fin}}}$.

An illustration of the system's behavior during a quench is shown in Figure~\ref{fig2}.
This leads to coarsening, where domains form and grow as $t^{1/2}$~\cite{Bray1994}.
The system thermalizes only on timescales $t \propto N^2 = 10^4$.
Consequently, the system remains in the high-NSR regime during observation.
Further details for the non-specialist reader can be found in Appendix~\ref{sec:app1}.

\section{Renormalization group for data analysis and GSA}\label{SDRG}

In this section, we review the FRG framework for signal detection in nearly continuous spectra as proposed in~\cite{RG1} and provide the essential details.
The interested reader can refer to our most recent paper~\cite{RG7}, the review~\cite{RG5}, and the forthcoming review~\cite{RG8}.
Additional details on the FRG formalism in field theory are available in~\cite{Delamotte}.

\subsection{General idea}

As discussed in the introduction, the primary challenge in anomaly detection for nearly continuous spectra lies in distinguishing the signal from the noise by identifying the information-bearing components.
The key idea developed in~\cite{RG1,RG2,RG3,RG4,RG5,RG6,RG7} is that this problem can be reduced to the renormalization group (RG) study of a specific statistical field theory.
In this framework, the field is analogous to the one in $\phi^4_d$ theory for the Ising model near the critical point, which captures the relevant long-range correlations between macroscopic regions of the ferromagnet (essentially 2-point and 4-point correlations).
Our approach follows the modern perspective on field theory established since the discovery of the RG~\cite{Zinn2}.
We construct the effective field theory through statistical inference as a maximum entropy model~\cite{Jaynes1}, which represents the least biased distribution consistent with the data.
This theory describes an unconventional type of scalar matter filling an abstract one-dimensional space.

Consistent with the effective field theory for the critical Ising model, the definition of this theory is independent of the specific data structure as long as the distribution of noisy degrees of freedom remains close to certain universal laws of random matrices, such as the MP distribution~\eqref{MP}.
This observation regarding the coarse-grained nature of the microscopic degrees of freedom is not an additional assumption but reflects the fundamental meaning of universality.
We identify the Gaussian fixed point of this field theory with the universal MP distribution.

The most elementary level of RG analysis occurs around the Gaussian fixed point through the distinction between relevant and irrelevant interactions via power counting.
It is at this level that GSA operates.

\subsection{Field theory and power counting}

Formally, we consider a local Euclidean field theory at equilibrium for a macroscopic field $\Psi: \mathcal{X} \to \mathbb{R}$.
This field is both a macroscopic observable and a 1-particle-irreducible (1PI) order parameter coupled to an external source $j: \mathcal{X} \to \mathbb{R}$.
The set $\mathcal{X} = \{0, \pm |p_1|, \ldots, \pm |p_{N_c}|\}$ contains $N_c \leq N$ eigenvalues from the bulk.
We define the generalized momentum $p_\mu$ from the correlation matrix eigenvalues $\lambda_\mu$ via the inference relation
\begin{equation}
    \frac{\delta^2 W_k}{\delta j(p_\mu) \delta j(-p_\mu)} = \lambda_\mu,
    \label{inference}
\end{equation}
where the free energy $W_k[j]$ defines the macroscopic field $\Psi(p_\mu) = \delta W_k / \delta j(-p_\mu)$.
This relation induces a distribution $\rho(p^2)$ on $\mathcal{X}$ from the eigenvalue distribution $\mu(\lambda)$.
In the continuous limit $N_c \to \infty$, for any sufficiently smooth function $f(x)$,
\begin{equation}
    \frac{1}{N_c}\sum_\mu f(p_\mu^2) \to \int \dd p^2 \,\rho(p^2) f(p^2).
    \label{continuum}
\end{equation}
The free energy is expressed as a path integral involving the model Hamiltonian and a scale-dependent mass term $R_k(p_\mu^2)$, or regulator, where $k \in \mathbb{R}^+$.
This regulator ensures that fluctuations are integrated out only up to the generalized momentum scale $k$.
The effective average action (EAA), $\Gamma_k[\Psi]$, is a modified Legendre transform
\begin{equation}
    \Gamma_k[\Psi] = \sup_{j} \left( \sum_{p \in \mathcal{X}} j(-p)\Psi(p) - W_k[j] \right) - \Delta H_k[\Psi],
\end{equation}
where the mass term $\Delta H_k[\Psi]$ is related to the regulator $R_k$ by
\begin{equation}
    \Delta H_k[\Psi]=\frac{1}{2} \sum_{p \in \mathcal{X}} \Psi(p) R_k(p^2) \Psi(-p).
\end{equation}
The regulator $R_k$ vanishes as $k \to 0$, such that all fluctuations are integrated out and the EAA reduces to the standard effective action.
The evolution of the couplings within $\Gamma_k[\Psi]$ as fluctuations are integrated is described by the Wetterich-Morris equation~\cite{Delamotte,Wett},
\begin{equation}
    \dot{\Gamma}_k=\frac{1}{2}\sum_{p \in \mathcal{X}} \dot{R}_k(p^2) \Big(\Gamma^{(2)}_k+R_k\Big)^{-1}(p,-p),
    \label{Wett}
\end{equation}
where $\Gamma^{(2n)}_k$ denotes the $2n$-th functional derivative of $\Gamma_k$ with respect to the classical field and the dot signifies the derivative with respect to $t = \ln k$.
The term $\Gamma^{(2)}_k + R_k$ is the inverse of the two-point function $G(p^2) = \delta^2 W_k / \delta j(p_\mu) \delta j(-p_\mu)$.
The inference condition~\eqref{inference} implies
\begin{equation}
    \Gamma_{k=0}^{(2)}(p=0) \equiv \lambda_0^{-1},
\end{equation}
where $\lambda_0 = \lambda_{\text{max}} - \lambda_{\text{min}}$ is the spectral width of the bulk and $\lambda_0^{-1}$ represents the effective mass.

\begin{figure}
    \begin{center}
        \includegraphics[width=0.55\textwidth]{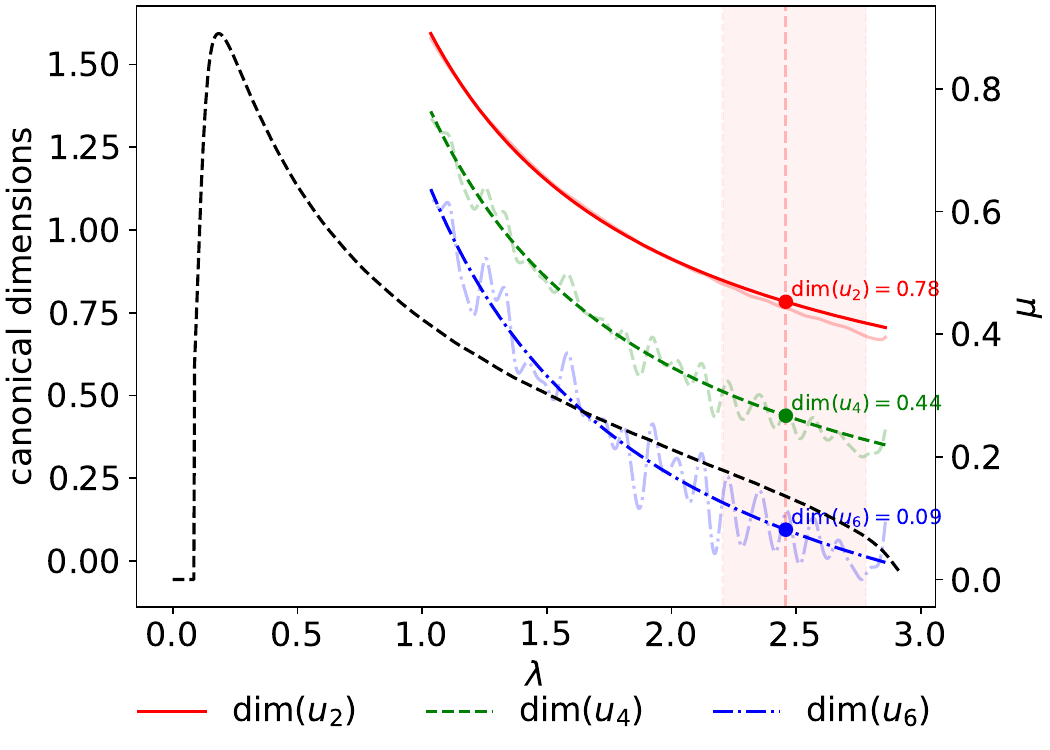}
    \end{center}
    \caption{%
        Canonical dimensions for a random matrix $X \in \mathbb{R}^{N \times P}$, where $N = 4 \times 10^4$, and $P / N = 0.5$, plotted against the interpolation of the eigenvalue spectrum.
    }\label{fig4}
\end{figure}

Although~\eqref{Wett} is exact, it cannot be solved analytically.
We employ the Local Potential Approximation (LPA), justified by power counting and numerical analysis~\cite{RG1},
\begin{equation}
    \Gamma_k[\Psi] = \frac{1}{2}\sum_{p \in \mathcal{X}} \Psi(p) (p^2+u_2(k)) \Psi(-p) + {U}_k[\Psi],
\end{equation}
where the potential is local,
\begin{equation}
    {U}_k[\Psi] = N_c\sum_{n=2}^\infty \frac{u_{2n}}{(2n)! N_c^n}  \sum_{\{p_i\}} \delta \left(\sum_{j=1}^{2n} p_j\right) \prod_{i=1}^{2n} \Psi(p_i),
\end{equation}
with the powers of $N_c$ adjusted for consistency in the continuous limit $N_c \to \infty$.
In the strict LPA, the derivative expansion neglects powers of $p^2$ beyond the leading Gaussian term, and the classical field reduces to its zero-mode component $\Psi(p) \approx \Psi \delta_{0p}$~\cite{Delamotte}.
The flow equation for $\mathcal{U}_k[\Psi]:=\Gamma_k[\Psi \delta_{0p}]$ then follows from~\eqref{Wett}.
Using the Litim regulator~\cite{Litim}, $R_k(p^2) = (k^2-p^2)\theta(k^2-p^2)$, we obtain the continuous limit
\begin{equation}
    \dot{\mathcal{V}}_k[\chi]=\frac{1}{2}\int dp^2 \rho(p^2) \dot{R}_k(p^2) \frac{k^2}{k^2 + \mathcal{V}_k^\prime(\chi)+2\chi \mathcal{V}^{\prime\prime}(\chi)},
    \label{floweqV}
\end{equation}
where $N_c\chi = \Psi^2/2$ and $\mathcal{V}_k(\chi) = \mathcal{U}_k[\Psi]$.
In the symmetric phase, the two-point function is
\begin{equation}
    G(p^2) = \frac{1}{p^2_\mu + m_{\text{eff}}^2}.
\end{equation}
The inference relation~\eqref{inference} then implies $m_{\text{eff}}^2 = \lambda_0^{-1}$ and
\begin{equation}
    \rho(p^2) = \frac{1}{(p^2 + \lambda_0^{-1})^2} \mu \left( \frac{1}{(p^2 + \lambda_0^{-1})} + \lambda_{\text{min}} \right).
\end{equation}
The flow equations for the couplings $u_{2n}$ involve the integral
\begin{equation}
    L(k) = \int_0^k \dd p\, p\, \rho(p^2).
\end{equation}
We redefine the couplings $u_{2n} \to {\bar{u}}_{2n}$ to remove the explicit $k$-dependence, shifting it into a linear term whose coefficient defines the canonical dimension,
\begin{equation}
    \frac{\dd}{\dd \tau}\bar{u}_{2n}= - \mathrm{dim}_\tau(u_{2n}){\bar{u}}_{2n} + \mathcal{O}({\bar{u}}_{2m}{\bar{u}}_{2q}),
\end{equation}
with
\begin{equation}
    \mathrm{dim}_\tau(u_4)=-2\left(\frac{t^{\prime\prime}}{t^\prime}+t^\prime\left(\frac{1}{2} \frac{\dd \ln{\rho}}{\dd t}-1\right)\right),
\end{equation}
\begin{equation}
    \mathrm{dim}_\tau(u_{2n})=-2(n-2) \frac{\dd t}{\dd \tau}+(n-1) \mathrm{dim}_\tau(u_4),
\end{equation}
where $\dd \tau = \dd \ln L(k)$ and $t^\prime = \dd t/ \dd \tau$.
The sign of the canonical dimension determines the relevance of the couplings: couplings with $\mathrm{dim}_\tau(u_{2n}) \geq 0$ are relevant, while those with $\mathrm{dim}_\tau(u_{2n}) < 0$ are irrelevant.
Unlike standard field theories, the canonical dimension here is scale-dependent\footnote{%
In ordinary field theories on $\mathbb{R}^d$, the distribution follows a power law $\rho(p^2) \sim (p^2)^{\frac{d-2}{2}}$.
Thus, $L(k)$ is a power of $k$, and the canonical dimensions are simple numbers, which are furthermore fixed by dimensional analysis related to the existence of a background structure.
In the case at hand, there is no external notion of scale.
The dimension is simply determined by the behavior of the flow around the Gaussian fixed point.
For more details, see~\cite{RG1} or~\cite{LahocheBeyond,lahoche2024functional,achitouv2024time,natta2024ward,achitouv2025constructing} in other contexts.
}.
Figure~\ref{fig4} illustrates the behavior for an analytical MP distribution and an i.i.d.\ Gaussian matrix $X$.
The MP universality class is characterized by two relevant couplings in the deep infrared (IR) limit, corresponding to the tail of the spectrum where $p_\mu^2 \ll 1$.
In this region, where the signal is presumed to reside, the quartic coupling is relevant and the sextic coupling is marginal.
Asymptotically, for an MP distribution, $\mathrm{dim}_\tau(u_{2n}) \propto (3-n)$.

\subsection{Generalized scale analysis}

\begin{figure*}[t]
    \centering
    \includegraphics[width=0.48\textwidth]{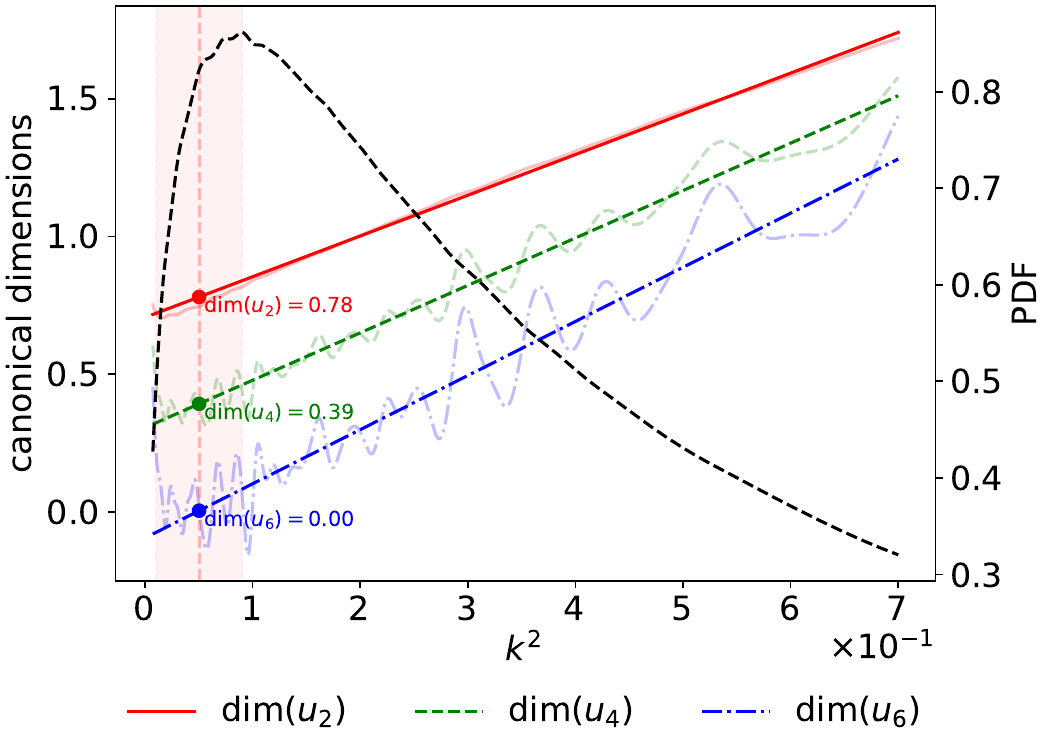} \hfil
    \includegraphics[width=0.48\textwidth]{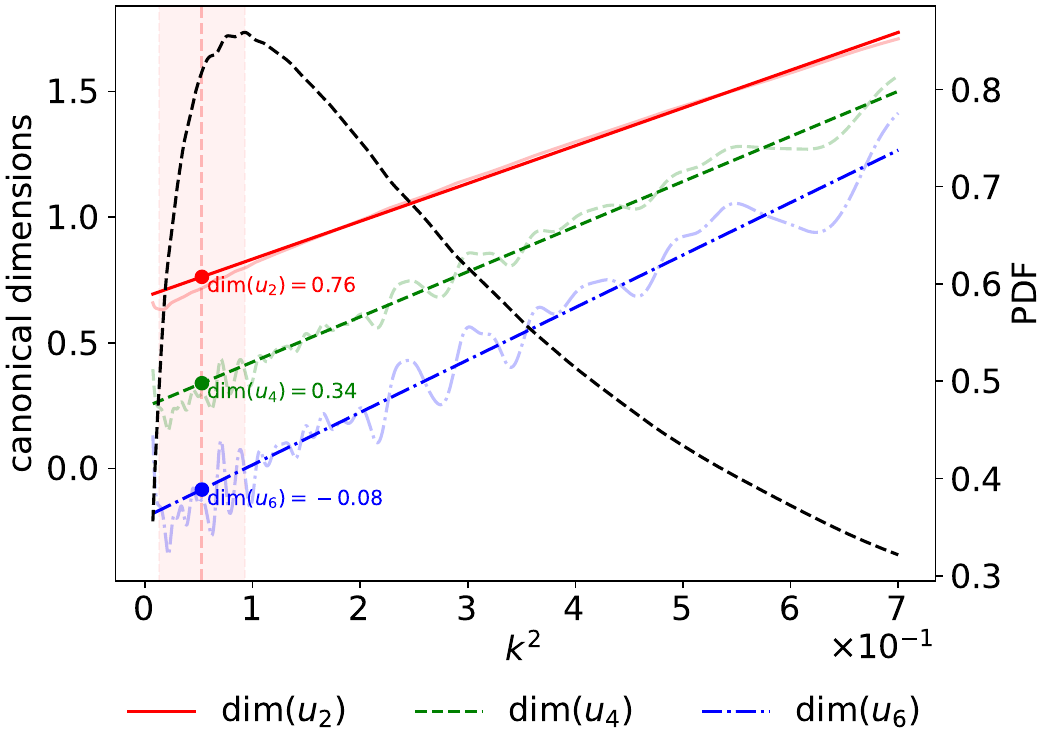}\hfil
    \includegraphics[width=0.48\textwidth]{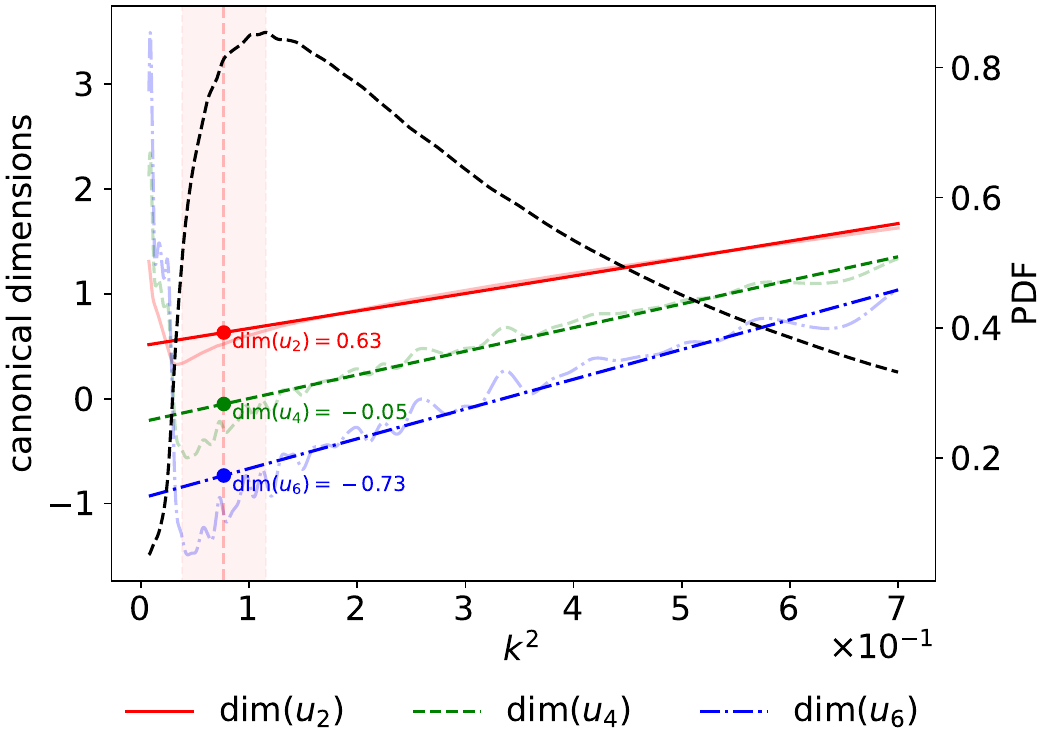}
    \caption{%
        Behavior of the canonical dimension by decreasing the NSR from a pure MP contribution (top left), to intermediate values (top right), to lower values (bottom). Data extracted from our previous work~\cite{RG7}.
    }\label{fig5}
\end{figure*}

Initially introduced in~\cite{RG1}, GSA identifies the non-equilibrium phase boundary $\Lambda$ (see Figure~\ref{fig1}) by tracking the spectrum's ability to reduce the dimension of the theory's renormalizable coupling subspace.
In general, it has been numerically demonstrated using model data with adjustable NSR that: (i) when the NSR is very high, the behavior of the canonical dimensions remains close to the Gaussian fixed point predicted by the MP distribution, where only $u_4$ is relevant; and (ii) decreasing the NSR triggers a transition to an ordered phase where renormalizable couplings become irrelevant, such that $\mathrm{dim}(u_{4}) < 0$.
Figure~\ref{fig5} illustrates this observation, and we refer the reader to our previous study~\cite{RG7} for a discussion on the influence of finite-$N_c$ effects.
For reproducibility, code can be found on \href{https://github.com/thesfinox/frg-signal-detection}{https://github.com/thesfinox/frg-signal-detection}.

The MP spectrum defines the disordered phase of the data-driven effective field theory, acting as a universal attractor for stochastic noise:
\begin{definition}
    Let $C$ be a Wishart\footnote{Let $X \in \mathbb{R}^{T \times N}$ whose entries are i.i.d.\ random variables following a centered normal distribution: $X_{ti} \sim \mathcal{N}(0, \sigma^2).$ The Wishart matrix $W$ is defined by the product: $W = T^{-1} X^T X.$} matrix depending on some parameters $(\alpha_1, \alpha_2, \ldots,\alpha_k)$ and let $I_k \subset \mathbb{R}^k$ the domain of the parameters. Then, $C$ is in the MP universality class at $p \in I_k$ if:
    \begin{enumerate}
        \item The empirical spectrum of $C$ converges toward the MP distribution as $P \to \infty$ with $q$ fixed.
        \item For any open set $U_p$ around $p$, there exists a constant $K > 0$ such that
              \begin{equation}
                  \operatorname{TVD}(C(p), C(q))
                  =
                  K \Vert p - q \Vert + \mathcal{O}\left(\Vert p - q \Vert^2\right).
              \end{equation}
    \end{enumerate}
    where $\operatorname{TVD}$ is the \emph{Total Variation Distance}:
    \begin{equation}
        \operatorname{TVD}(C(p),C(q))
        =
        \int \dd\lambda\, \left| \mu_p(\lambda)-\mu_q(\lambda) \right|,
    \end{equation}
    and $\Vert p-q \Vert$ is the induced Euclidean distance over $I_k$.
\end{definition}

Following our results in~\cite{RG7}, a pessimistic yet robust estimation of the signal-to-noise boundary $\Lambda$, denoted as $\Lambda_{\text{exp}}$, is defined as:
\begin{equation}
    \Lambda_{\text{exp}}
    =
    \min_\lambda (\lambda_{c,u_{4}}, \lambda_{c,u_{6}}),
\end{equation}
where $\lambda_{c,u_{2n}}=\lambda_{\mathrm{dim}(u_{2n})=0}$ is the point in the spectrum where the coupling becomes marginal.
Generally, $\lambda_{c,u_{4}} > \lambda_{c,u_{6}}$, but $\lambda_{c,u_{6}}$ is numerically unstable, whereas $\lambda_{c,u_{4}}$ benefits from an eigenvalue gap that is larger than the typical fluctuations associated with the intrinsic variability of the data ensemble, see~\cite{RG7}.
This crossing point marks the physical threshold where non-perturbative interactions overcome Gaussian fluctuations.
Therefore, in this paper, we choose $\Lambda_{\text{exp}} = \lambda_{c,u_{4}}$ in practice.

\section{Critical temperature estimation}\label{Critical}

\begin{figure*}[t] 
    \centering
    \includegraphics[width=0.48\textwidth]{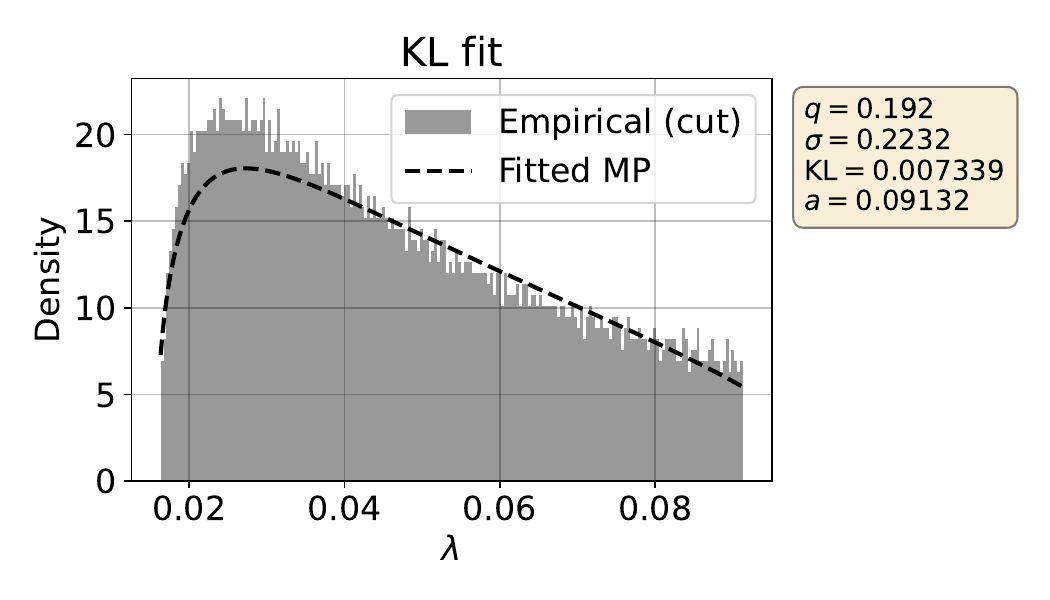}
    \hfil
    \includegraphics[width=0.48\textwidth]{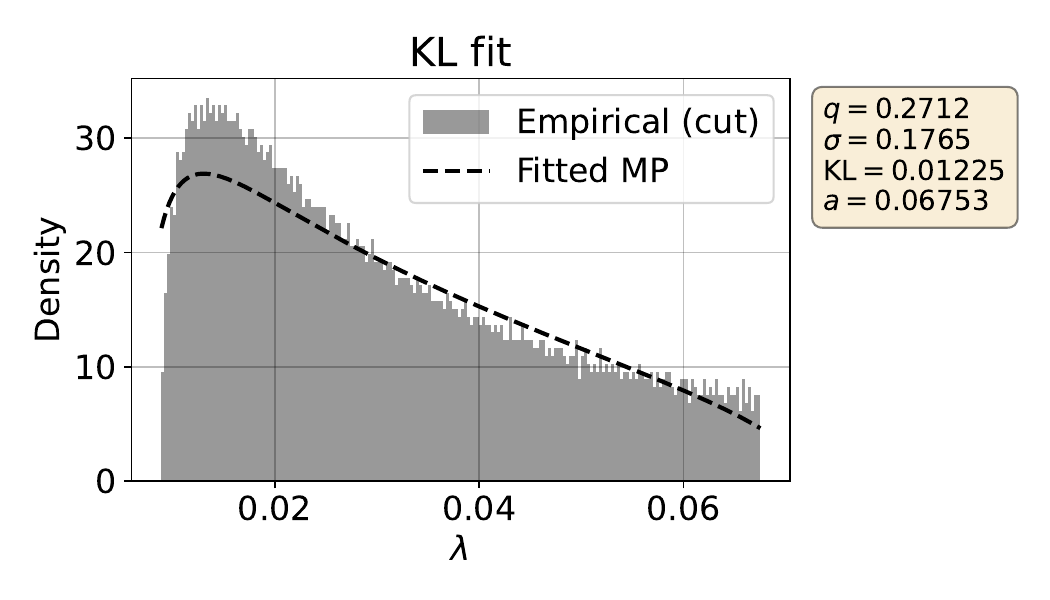}
    \hfil
    \includegraphics[width=0.48\textwidth]{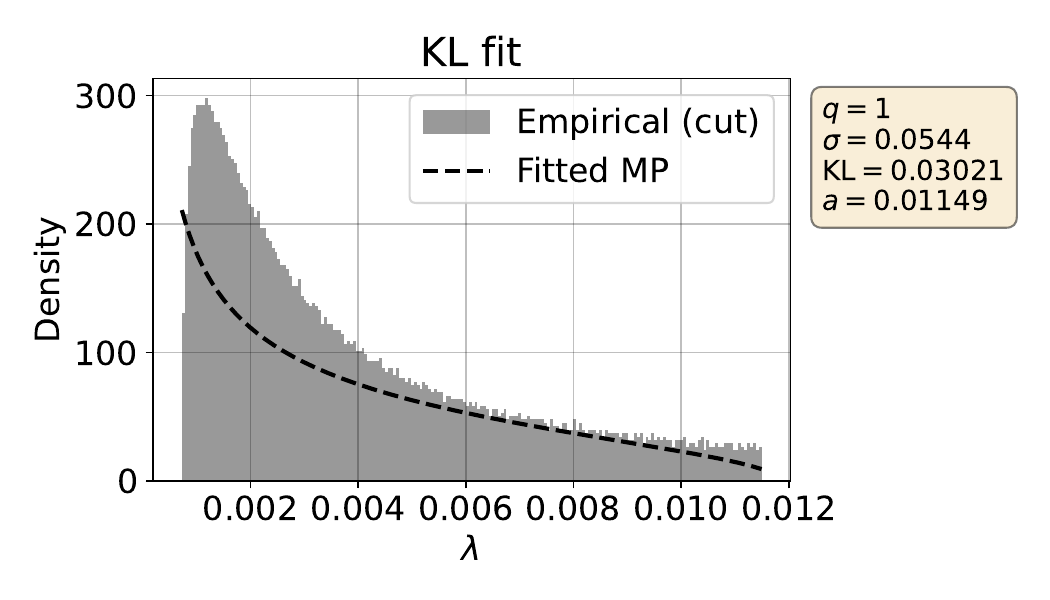}
    \caption{%
        Behavior of the spectrum found by the KL-proxy for high temperature, $T=5$ (top left), in the vicinity of the critical regime $T=0.648$ (top right) and below the critical temperature $T=0.350$ (bottom) for $b=0.8$.
    }\label{fig3}
\end{figure*}

The data is structured as a time series, represented by a matrix $X \in \mathbb{R}^{ t_{\text{fin}} \times \mathcal{N}}$, where $\mathcal{N} = N\times N$ denotes the total size of the 2D grid and $t_{\text{fin}} = 4 \times 10^4$ the number of time steps.
At high temperatures, the model is nearly Gaussian, and the spectrum of the $\mathcal{N} \times \mathcal{N}$ correlation matrix $C$ closely follows the distribution predicted by Theorem~\ref{th1} (Figure~\ref{fig3}, left).
Conversely, at lower temperatures, and particularly when crossing the critical point $T=T_c$, long-range correlations emerge and distort the spectrum, which then exhibits typical heavy-tailed behavior.
This phenomenon is linked to the appearance of low-mass IR modes, associated with long-range correlations and characteristic power-law scaling~\cite{vinayak2014spectral} (Figure~\ref{fig3}, right).

\begin{figure*}[t] 
    \centering
    \includegraphics[width=0.48\textwidth]{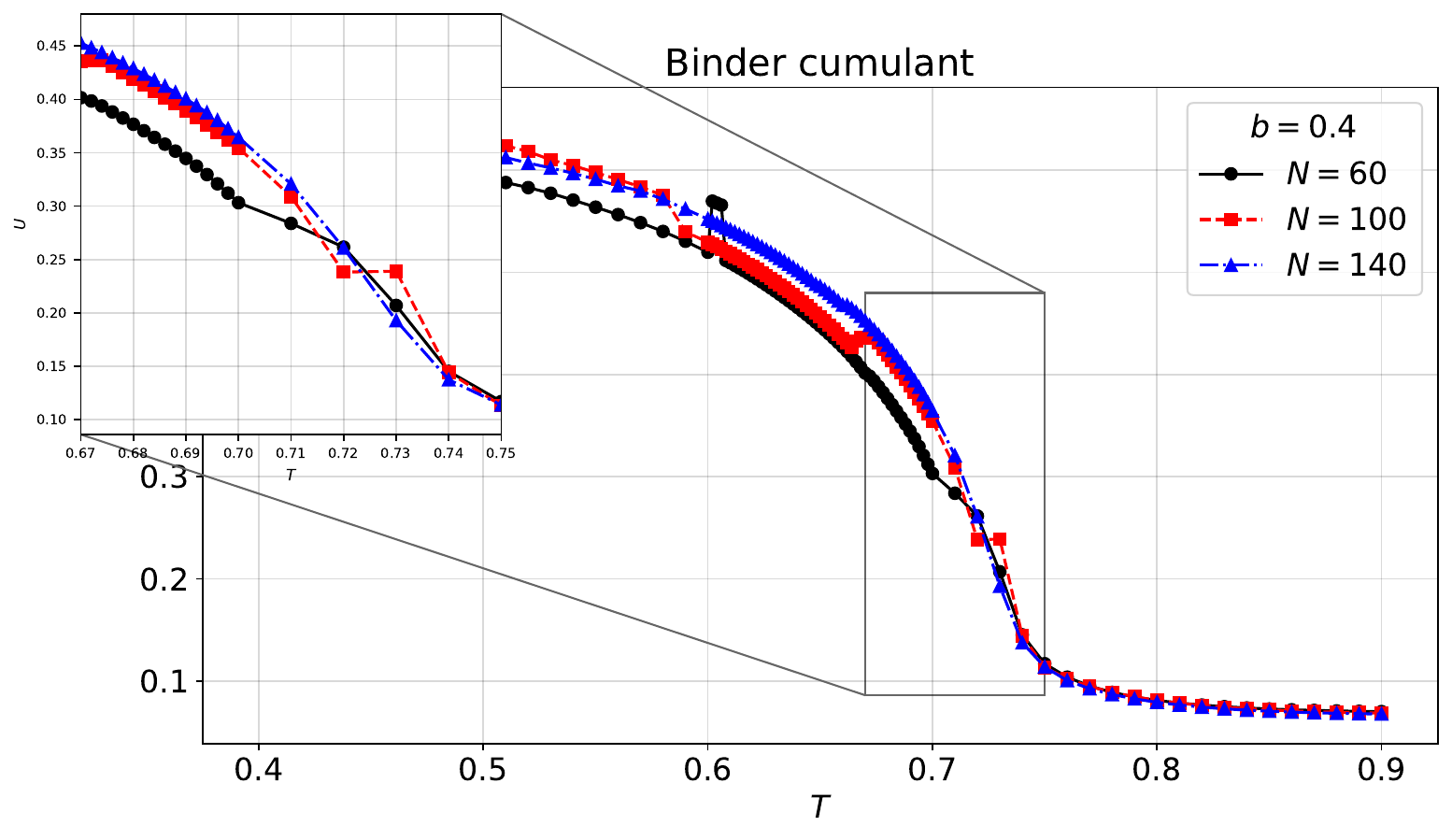}
    \hfil
    \includegraphics[width=0.48\textwidth]{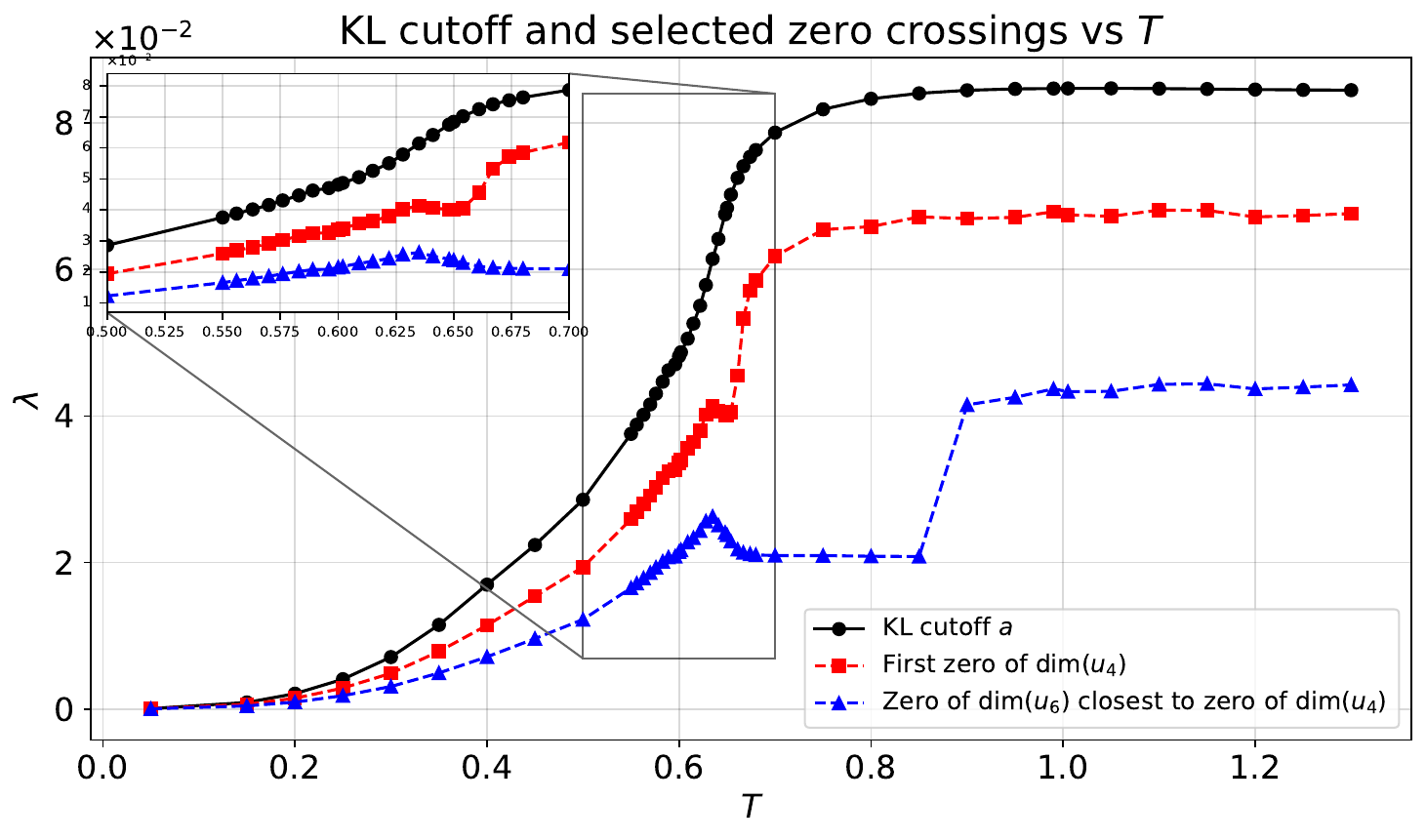}
    \caption{%
        Behavior of the Binder cumulant for different grid sizes with $b=0.4$ (left).
        Empirical behavior of $\lambda_c$ for different temperatures with different methods for $b=0.8$ (right).
    }  \label{fig6}
\end{figure*}

\paragraph{Model A} The critical temperature can be estimated using the Binder cumulants (BC)~\cite{selke2006critical}:
\begin{equation}
    K
    =
    1 - \frac{\langle \varphi^4 \rangle}{3 \langle \varphi^2 \rangle^2},
\end{equation}
interpolating between the high temperature where $K \approx 0$ (almost Gaussian regime) and the low temperature phase $K \approx 2/3$.
By varying the grid size, the physical transition temperature is identified as the crossing point of the different curves\footnote{%
    A complementary method consists of studying the behavior of the magnetic susceptibility.
    However, the identification of the critical temperature, while in agreement with the Binder cumulant method, would be less precise.
    Furthermore, this method exhibits a (numerical) singularity at low temperature, which essentially stems from two sources: the $1/T$ factor, and the fact that tunneling decouples from coarsening, i.e.\ the variance does not go to zero fast enough to compensate for the $1/T$ factor.
} (Figure~\ref{fig6}, left).
For each value of $b$, we thus obtain an experimental value $T_c^{(\text{exp})}(b)$ and we study the behavior of the eigenvalue distribution of the matrix $C$ across different temperatures.
The NSR boundary $\Lambda_{\text{exp}}(T)$ defined above also depends on temperature: a qualitative change in the distribution between high temperatures (Gaussian behavior) and low temperatures (appearance of large spin clusters) can be interpreted as the emergence of a signal.
The transition region, around $T_c$, thus marks the precise signal-to-noise transition point and is reflected in the empirical behavior of $\Lambda_{\text{exp}}(T)$ (see Figure~\ref{fig6}, right).
The transition point, indicated on the figure, corresponds to the inflection point, showing a discontinuity that becomes increasingly pronounced as the system size increases.
This point corresponds to the shift from a high-temperature regime characterized by a distribution very close to MP (according to Theorem~\ref{th1}), to a low-temperature regime where an increasing portion of the degrees of freedom is captured by the emerging order (such that $\Lambda_{\text{exp}}(T) \to 0$ as $T \to 0$).
At the transition point, a macroscopic fraction of the degrees of freedom becomes ordered and thus abruptly exits the noise-dominated regime\footnote{%
    Note that at high temperatures, the distribution plateaus, as the typical scale of fluctuations tends toward a constant.
}.

\begin{figure}
    \centering
    \includegraphics[width=0.7\textwidth]{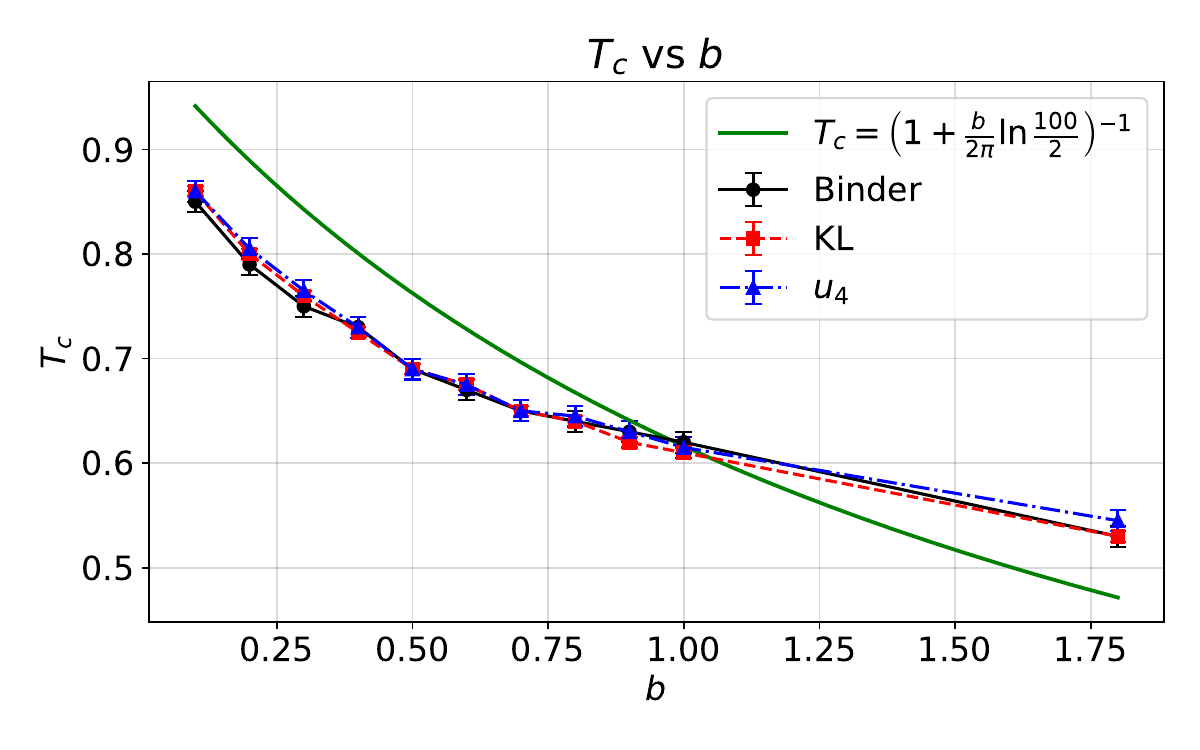}
    \caption{%
        Comparison of the critical temperatures $T_c(b)$ obtained by the various methods, for different values of $b$.
        The final point at $b \approx 1.8$, corresponding to the ``Ising-like'' case, is in good agreement with Onsager's temperature ($\approx 0.57$).
        The mean-field approximation obtained through the Hartree method is given by the green curve.
    }\label{fig7}
\end{figure}

We propose to compare our GSA-based estimation of $\lambda_c$ with a standard method, similar to the one considered by Bouchaud and Potters~\cite{Bouchaud1} by minimizing the Kullback-Leibler divergence,
\begin{equation}
    D_{KL}(\mu \parallel \nu) = \int_{-\infty}^{+\infty} \dd x \, \mu(x) \ln \left( \frac{\mu(x)}{\nu(x)} \right),
\end{equation}
between the empirical distribution and a MP distribution with respect to the parameters $(\sigma^2, q)$ (see Theorem~\ref{th1}).
The behavior of $\lambda_c(T)$ is shown in Figure~\ref{fig6} (bottom) and is found to be very close to that derived from the GSA.
In the same manner, we extract a measurement of the critical temperature $T_c^{(KL)}(b)$ for different values of $b$.
Figure~\ref{fig7} summarizes our findings, comparing the three temperatures: $T_c^{(exp)}(b)$, $T_c^{(GSA)}(b)$, and $T_c^{(KL)}(b)$.

\begin{figure*}[t] 
    \centering
    \includegraphics[width=0.48\textwidth]{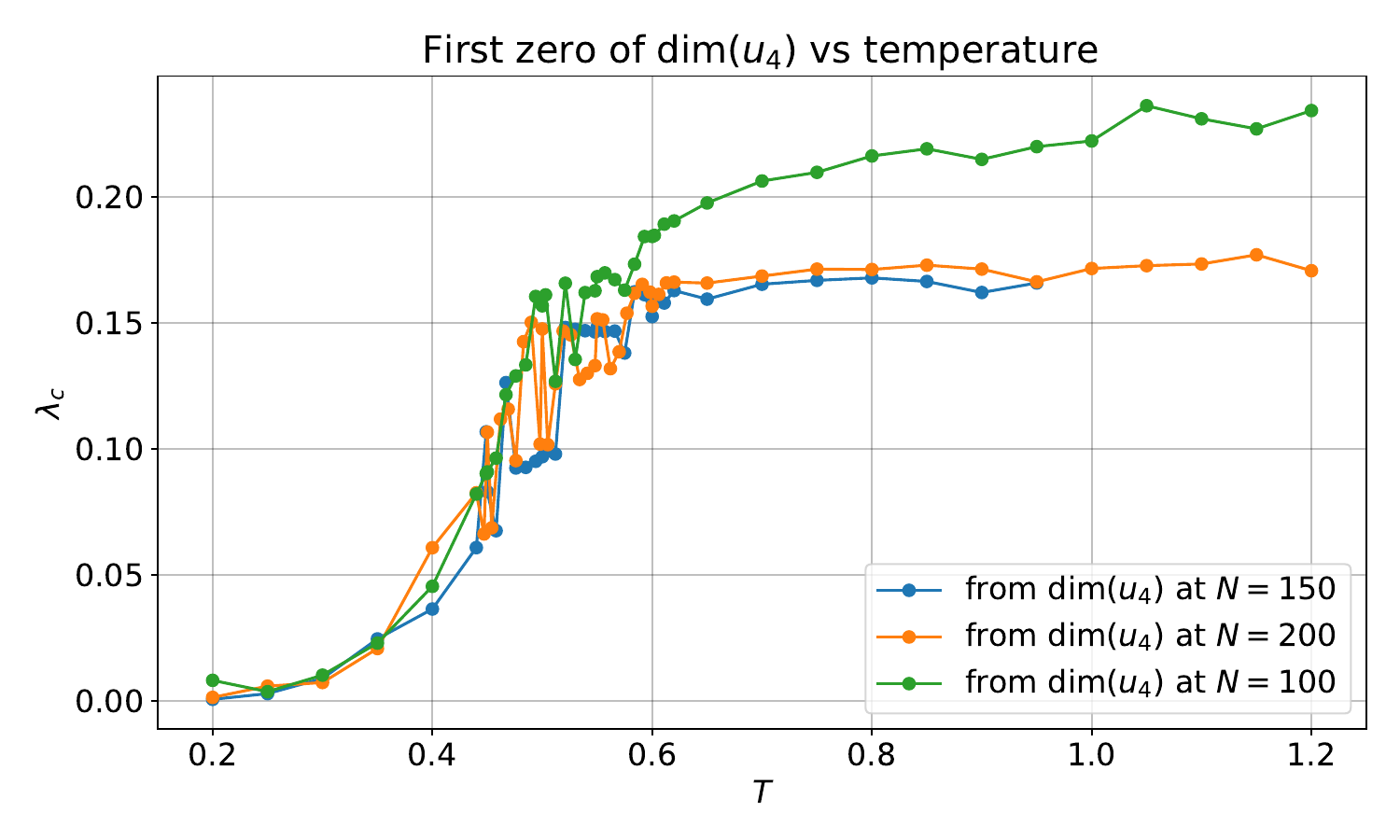}
    \hfil
    \includegraphics[width=0.48\textwidth]{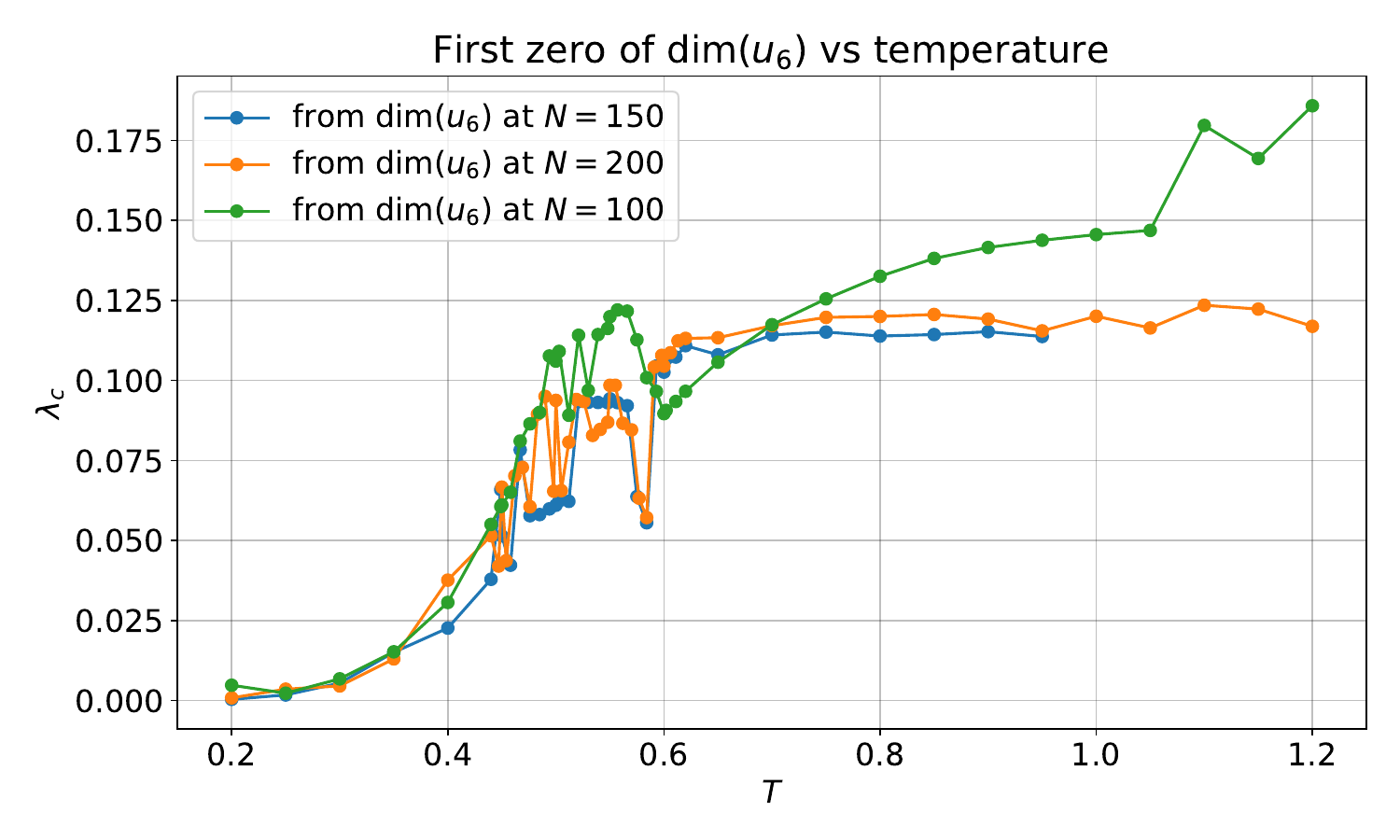}
    \hfil
    \includegraphics[width=0.48\textwidth]{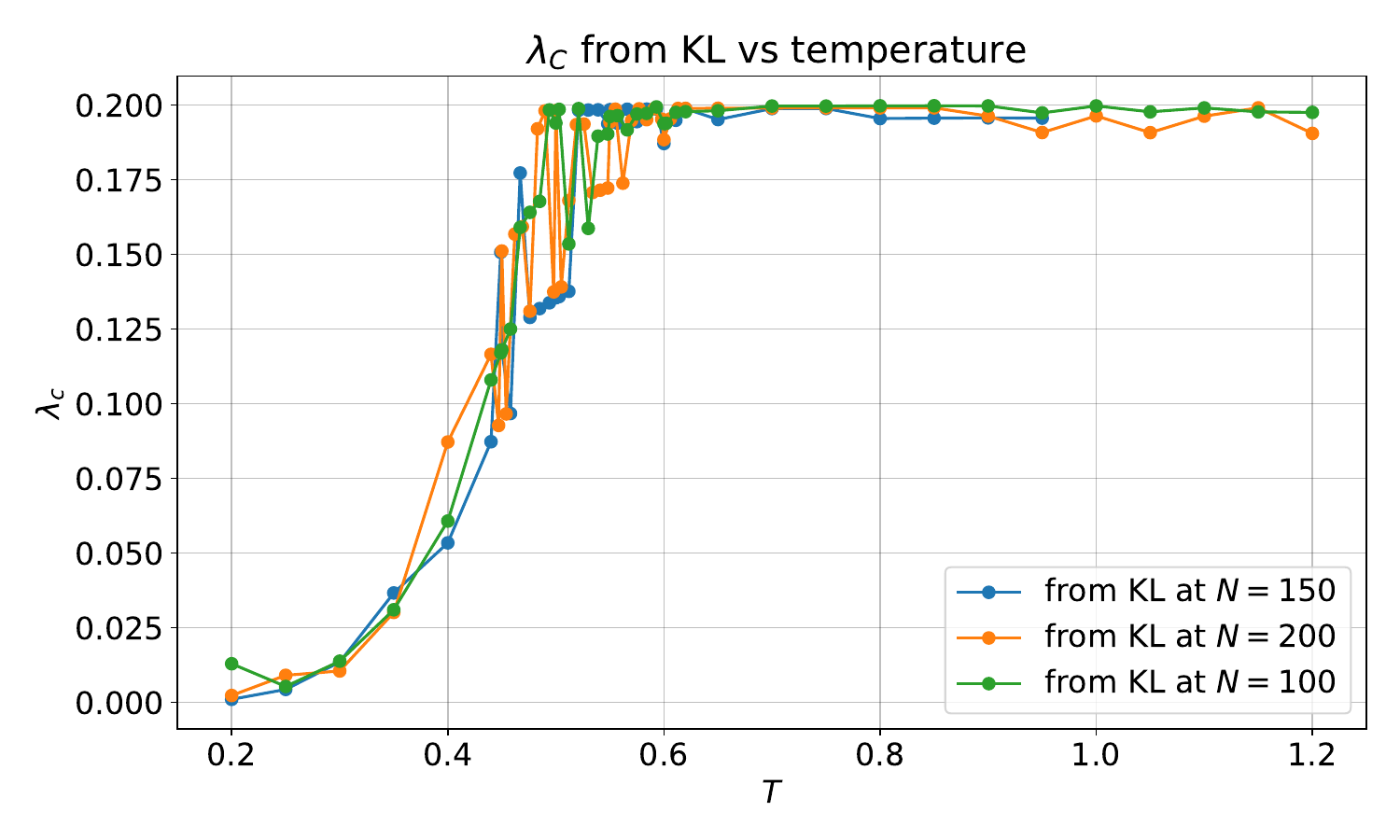}
    \caption{%
        Position of $\lambda_c$ as a function of $T$ with different methods: $\dim(u_4)$ (top left), $\dim(u_6)$ (top right), and KL divergence (bottom).
    }\label{fig44}
\end{figure*}

\begin{figure*}[t] 
    \centering
    \includegraphics[width=0.48\textwidth]{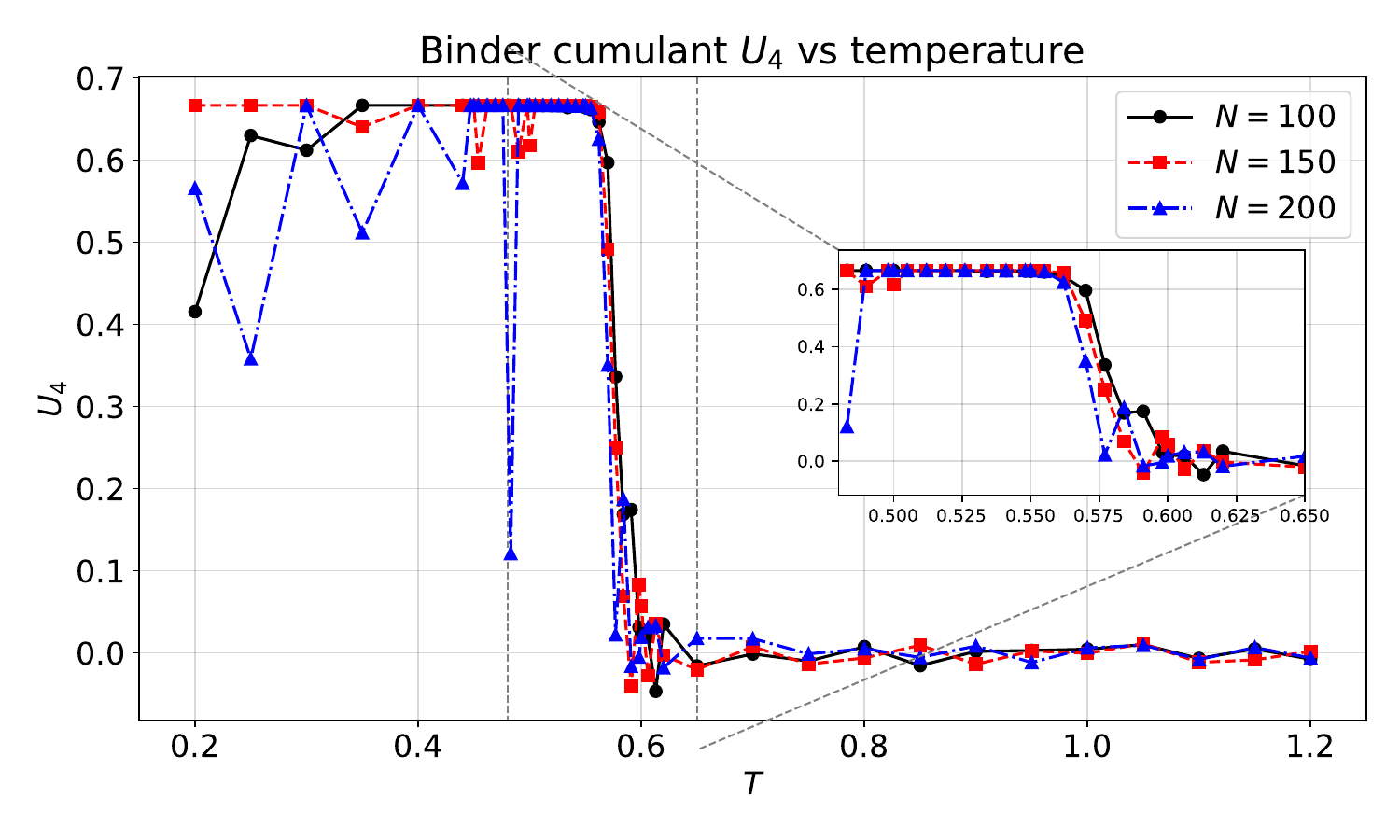}
    \hfil
    \includegraphics[width=0.48\textwidth]{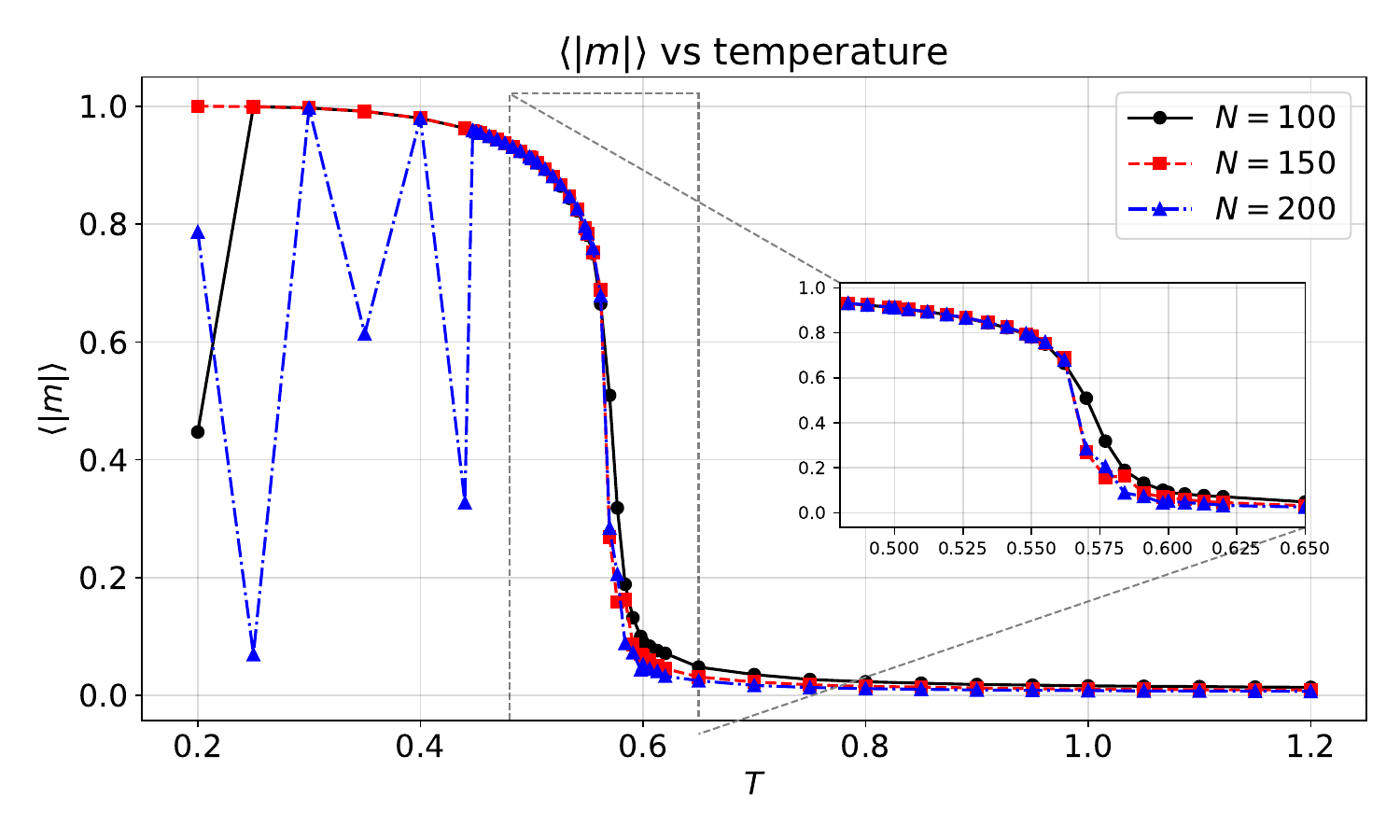}
    \caption{%
        Binder cumulant (left) and absolute magnetization (right) for the Ising model.
    } \label{fig45}
\end{figure*}

\paragraph{Ising model} As a complement, let us analyze the same behavior for the pure Ising model, corresponding to the parameterization in \eqref{VIsing} as $\varepsilon\to \infty$, using a Monte Carlo-type method with $J=1/4$.
The results are summarized in Figures~\ref{fig44}~and~\ref{fig45}.
The Binder cumulants method provides an estimate of the critical temperature $T_c \in [0.58, 0.6]$, which represents a deviation of less than 2\% compared to the analytical prediction by Onsager ($T_c \approx 0.567$ for $J=1/4$).
The magnetization method provides a similar result, in the optimistic window $T_c \approx [0.58, 0.6]$.

The figures, which are comparable to those obtained for Model A, clearly show three distinct regimes:
\begin{enumerate}
    \item A high-temperature regime $T > 0.6$, where the curves form a plateau. This is a consequence of the chosen parameterization of the potential: as $T \to \infty$, the field effectively tends toward a large-mass Gaussian regime where fluctuations are suppressed.
    \item A transition regime $0.45 \le T < T_c$ marked by strong instability: the correlation length $\xi$ becomes so large that the noise-signal separation becomes numerically unstable.
    \item A low-temperature regime $T < 0.45$, where the cutoff $\Lambda$ decreases because the signal (order) occupies the entire space and the NSR becomes low, with the majority of available degrees of freedom being captured by the emerging order.
\end{enumerate}
The presence of oscillations and instability in the critical region is numerically expected.
It is primarily a consequence of the divergence of the correlation length and critical slowing down.
This critical slowing down causes instabilities to persist, as they are no longer suppressed over long time scales, making the simulation unstable: due to strong correlations, the flip of a single spin is very likely to be rejected by the algorithm.
This is essentially due to the fact that at $T_c$, the system can no longer distinguish between thermal fluctuations (noise) and domain fluctuations (signal), because the correlation length $\xi$ is infinite.

It is noteworthy that the transition is significantly clearer via GSA, specifically through the quartic dimension, but even more so through the sextic dimension: the transition toward the high-temperature plateau (where the sextic coupling reaches its marginality threshold and MP noise dominates) occurs around $T = 0.58 \pm 0.01$, which aligns perfectly with the theoretical prediction.
This constitutes yet another rigorous validation of GSA: the canonical dimension of the sextic coupling detects the Ising transition with less than 2\% error, simply by observing when the interaction couplings reach the marginality threshold.
Note that the KL method predicts a lower value, around 0.53, corresponding to a deviation of approximately 10\%.
This suggests that the KL method is significantly more impacted by the discrete nature of the Ising transition than the GSA approach.

As a final remark, notice that numerical instabilities appear stronger in this case than in Model A.
This is primarily caused, in the Ising model, by the fact that the system can only overcome energy barriers through discrete jumps.
In contrast, the continuous nature of the Langevin dynamics in Model A provides a smoother evolution than the discrete Monte Carlo spin-flip dynamics.
For Model A, the results from the KL method were comparable to those from GSA.
However, the results presented here demonstrate that GSA not only reproduces but even improves the prediction of a known physical behavior by using signal detection as the underlying mechanism.
Finally, it is worth noting that the asymptotic scale (for high temperature) of the cutoff is of particular interest.
Based on the canonical dimension of the quartic coupling, the GSA cutoff value remains below the one provided by the minimization of the KL method.

\section{Discussion and Conclusion}\label{discuss}

In this paper, we establish a rigorous mapping between signal detection in quasi-continuous spectra, specifically at high NSR, and the determination of the critical temperature of a well-known 2D non-equilibrium model, namely, Model A.
By framing these transitions as an anomaly detection task and using GSA to perform the detection, we have shown that the signal-to-noise boundary not only matches experimental observations and the Onsager temperature, but also provides higher precision than standard estimation based on the KL divergence minimization.
These results establish GSA as a robust, high-precision approach for anomaly detection in the high-NSR regime.

GSA can be viewed as a minimal approach, as it only considers the flow behavior around the Gaussian fixed point.
In this specific case, it is indeed the only expected fixed point, as the $k$-dependence of the canonical dimensions precludes the existence of global non-Gaussian fixed points.
There are, however, fixed directions that lead to asymptotic fixed points, comparable to the ordinary Wilson-Fisher fixed point~\cite{RG7}.
The presence of a signal progressively drives the distance between this fixed point and the Gaussian point toward zero, thereby reducing the size of the region where the $\mathbb{Z}_2$ reflection symmetry is broken.
We will discuss the non-Gaussian behavior of the flow in a forthcoming paper~\cite{RG8}.

An interesting aspect of the physics of non-equilibrium systems is that we have mapped the measurement of the critical temperature of an interacting non-equilibrium system to the study of an equilibrium theory in the vicinity of the Gaussian fixed point.
Establishing this correspondence provides new insights, and the method may prove relevant for studying both equilibrium and non-equilibrium statistical physics systems, particularly as an addition to the arsenal of numerical methods for evaluating critical temperatures.
A distinct advantage is that this method only requires varying the temperature rather than the system size. It could be especially pertinent for systems exhibiting a lack of self-averaging, notably those with quenched disorder.
We plan to examine which phase transitions can be elucidated by GSA in future work.

\section*{Acknowledgements}
The authors acknowledge support from the COMETA COST Action \href{https://www.cost.eu/actions/CA22130/}{CA22130}.
VL expresses special thanks to the swan of destiny.

\appendix

\section{Large-N Solution and Hartree Approximation}\label{sec:app1}

This appendix provides a baseline for data analysts by deriving the logarithmic shift of the critical temperature in 2D within the Hartree approximation~\cite{mulet2007langevin,livi2017nonequilibrium}.
This Gaussian reference captures the role of fluctuations in suppressing long-range order, though it remains insufficient to reach the exact Onsager limit.
Let $x \in \mathbb{R}^d$ and consider a family of $N$ fields $\phi_i(x,t)$, whose evolution follows a Langevin-type equation:
\begin{equation}
    \frac{\dd}{\dd t} \phi_i(x,t)
    =
    \Delta \phi_i(x,t)
    -
    2 N \phi_i(x,t) V^\prime(\phi^2(x,t))
    +
    \eta(x,t),
\end{equation}
where $\Delta$ is the standard Laplacian over $\mathbb{R}^d$, $\eta$ is a Gaussian white noise with zero mean and variance $\langle \eta(x,t) \eta(x^\prime,t^\prime) \rangle = 2 T \delta(x-x^\prime) \delta (t-t^\prime)$, and $\phi^2 = N^{-1} \sum_i \phi_i^2(x,t)$.
We shall use a quartic potential:
\begin{equation}
    V(\phi^2) = \frac{1}{2}a \phi^2 + \frac{1}{4} b (\phi^2)^2,
\end{equation}
with $a = (T-T_0)$ and $b \geq 0$.
When $b = 0$, the system becomes unstable at $T = T_0$, with the field diverging exponentially for $T < T_0$.
The introduction of a non-zero interaction ($b > 0$) stabilizes the dynamics, ensuring convergence toward the equilibrium states where $V^\prime = 0$.
However, the interaction also competes with the development of long-range order, associated with a diverging spatial correlation length, reducing the physical critical temperature to $T_c < T_0$.
Furthermore, when initialized at infinite temperature and suddenly quenched to $T < T_c$, the system exhibits an equilibration time that scales with $N$.

In the large-$N$ limit, the quantity $\phi^2$ is assumed to be sharply peaked around its mean $R^2$.
For sufficiently large $t$, $R$ can be considered time-independent.
The equation thus linearizes, and in Fourier space:
\begin{equation}
    \frac{\dd}{\dd t} \phi_i(k,t) = -k^2\phi_i(k,t) - (a + b R^2) \phi_i(k,t) + \eta(k,t).
\end{equation}
The critical temperature $T_c$ is now $T_c = T_0 - b R^2 \le T_0$.
The solution to the previous equation becomes:
\begin{equation}
    \phi_i(k,t)
    =
    \phi_i(k,0)e^{-(k^2+a_{\text{eff}}) t}
    +
    \int_0^t \dd t^\prime \eta(k,t^\prime) e^{-(k^2+a_{\text{eff}}) (t-t^\prime)}.
\end{equation}
Assuming purely random initial conditions $\phi_i(x,0) \sim \mathcal{N}(0,1)$ (high-temperature phase, $T \gg 1$), and that the system is suddenly quenched at $t=0$ to a temperature $T < T_c$, one can calculate the correlation of the Fourier modes after averaging over noise fluctuations:
\begin{equation}
    \langle \phi(k, t) \phi(k', t) \rangle
    =
    (2\pi)^d \delta(k+k')\, C(k, t),
\end{equation}
where
\begin{equation}
    C(k, t)
    =
    e^{-2(k^2+a_{\text{eff}})t} + T\, \frac{1 - e^{-2(k^2+a_{\text{eff}})t}}{k^2+a_{\text{eff}}}.
    \label{eq:correlationskk}
\end{equation}
For sufficiently large $t$, such that memory of the initial conditions is lost exponentially, $C(k, t) \approx (k^2+a_{\text{eff}})^{-1}$, leading to the self-consistent equation:
\begin{equation}
    a_{\text{eff}}
    =
    a + b T \int \frac{\dd^d k}{(2\pi)^d} \frac{1}{k^2+a_{\text{eff}}}.
\end{equation}
The critical temperature is precisely defined by $a_{\text{eff}} = 0$:
\begin{equation}
    0
    =
    a + b T_c \int \frac{\dd^d k}{(2\pi)^d} \frac{1}{k^2}
    =
    a + \frac{b T_c}{2\pi} \ln \left(\frac{M}{\mu}\right),
\end{equation}
where $M$ and $\mu$ are UV and IR cut-offs ($\mu \leq \vert k \vert \leq M$).
The Hartree approximation, which extends standard mean-field theory, consists of applying this self-consistent relation to finite $N$, specifically $N = 1$.
For a grid of size $L\times L$ ($L\in \mathbb{N}$) and step $\ell$, the first Brillouin region is $[-\pi/\ell, \pi/\ell]$, the UV cut-off is then $\pi/\ell$, and the IR is $2\pi/(L \ell)$.
Hence:
\begin{equation}
    T_c = \frac{T_0}{1 + \frac{b}{2\pi} \ln \left(\frac{L}{2}\right)}.
\end{equation}
The critical temperature vanishes in the thermodynamic limit, reflecting the breakdown of mean-field theory for $d \leq 4$.
However, the underlying physical mechanism remains valid: fluctuations hinder the development of long-range order, reducing $T_c$ from its bare value $T_0$.
Taking the Fourier transform of~\eqref{eq:correlationskk}, we find that the real-space correlations exhibit a spatial decay governed by $\exp \left(-\frac{\vert x-x^\prime \vert^2}{8 t} \right)$, with the correlation length growing as $t^{1/2}$.
The system reaches equilibrium only on timescales proportional to the number of grid sites.
The size of the spin domains typically grows as $t^{1/2}$.

\section{Recovering the Ising Model}\label{sec:app2}

In order to show the correspondence with the Ising model, let us consider the discrete Laplacian~\eqref{DiscreteLaplacian}.
Consider moreover the discrete gradient $(\nabla_{\text{dis}} \varphi)_{ij}$ with components:
\begin{equation}
    (\nabla_{\text{dis}} \varphi)_{ij}
    =
    \begin{pmatrix}
        \varphi_{i+1,j}-\varphi_{ij} \\
        \varphi_{i,j+1}-\varphi_{ij}
    \end{pmatrix}
\end{equation}
Consider the quantity:
\begin{equation}
    I[\varphi]
    =
    \frac{1}{2}\, \sum_{i,j} [(\varphi_{i+1,j}-\varphi_{ij})^2+(\varphi_{i,j+1}-\varphi_{ij})^2],
\end{equation}
such that
\begin{equation}
    \frac{\partial I}{\partial \varphi_{ij}}=-(\Delta_{\text{dis}}\varphi)_{ij}.
\end{equation}
Let us assume the discrete kinetics:
\begin{align}
    \varphi_{i,j}^{n+1}
    =
    \varphi_{i,j}^{n}
    +
    \Delta t\left[\frac{J}{T} (\Delta_{\text{dis}}\varphi)^{n}_{i,j} - V'(\varphi_{i,j}^{n})\right]+\sqrt{2\Delta t}\,\xi_{i,j}^{n}\,
\end{align}
where $J>0$ and $\xi_{i,j}^{n} \sim \mathcal{N}(0,1)$.
Choose the potential $V$ in~\eqref{VIsing}, for $\varepsilon\to \infty$.
In the time continuum limit, the corresponding equilibrium path integral is
\begin{equation}
    Z^{(\text{eq})}_{\Delta t\to 0}
    =
    \int \prod_{i,j}\dd \varphi_{ij}\,
    e^{-\frac{J}{T}I[\varphi]}\; e^{-V[\varphi]}.
\end{equation}
The potential $V[\varphi]$ can be expanded around its extrema:
\begin{equation}
    V[\varphi]=V[\varphi_0]+\frac{1}{2} V^{\prime\prime}[\varphi_0] (\varphi-\varphi_0)^2+\mathcal{O}( (\varphi-\varphi_0)^3).
\end{equation}
Two solutions exists, $\varphi_0=\pm 1$, but $V[1] = V[-1] = -\varepsilon/2$ and $V^{\prime\prime}[1] = V^{\prime\prime}[-1]=2\varepsilon$.
Hence, in the $\varepsilon \to \infty$ limit, $e^{-V[\varphi]}$ behaves as a Dirac delta, enforcing $\varphi_{ij}=\pm 1$ in the path integral, namely,
\begin{equation}
    Z^{(\text{eq})}_{\Delta t\to 0} \approx \sum_{\{\varphi_{ij}=\pm 1\}}\, e^{-\frac{J}{T}I[\varphi]}.
\end{equation}
Now, we have:
\begin{equation}
    I[\varphi]
    =
    2 N^2 - \sum_{\langle I,J \rangle} \, \varphi_{I}\varphi_{J},
\end{equation}
where $I,J$ denote pairs of indices $I=(i,j)$ on the grid, and $\langle I,J \rangle$ means we sum only over nearest neighboring sites.
The partition function then becomes:
\begin{equation}
    Z^{(\text{eq})}_{\Delta t\to 0}
    =
    e^{-\frac{2 N^2J}{T}}\; \sum_{\{\varphi_{I}=\pm 1\}}\, e^{\frac{J}{T}\sum_{\langle I,J \rangle}\, \varphi_{I}\varphi_{J}}.
\end{equation}
which is, up to an irrelevant factor, the standard 2D Ising model.

\section*{References}
\printbibliography[heading=none]

\end{document}